# Stone structures in the Syrian Desert

Amelia Carolina Sparavigna
Dipartimento di Fisica,
Politecnico di Torino, Torino, Italy

An arid land, known as the Syrian Desert, is covering a large part of the Middle East. In the past, this harsh environment, characterized by huge lava fields, the "harraat", was considered as a barrier between Levant and Mesopotamia. When we observe this desert from space, we discover that it is crossed by some stone structures, the "desert kites", which were the Neolithic traps for the game. Several stone circles are visible too, as many Stonehenge sites dispersed in the desert landscape.

The Syrian Desert is an arid land of south-western Asia, extending from the northern Arabian Peninsula to the eastern Jordan, southern Syria, and western Iraq, largely covered by lava fields. Considered in the past as a barrier between Levant and Mesopotamia, it is now crossed by several routes and pipelines. This desert possesses two volcanic regions. One is the Jabal al-Druze, in the As-Suwayda Governorate. This volcanic field consists of a group of several basaltic volcanoes active from the lower-Pleistocene to the Holocene [1]. The other field is that of the Harrat Ash Shaam (in Arabic, the lava fields are the harraat, sing. harrah; before a name, harrat), including the Es-Safa volcano, in the South of Syria, south-east of Damascus. According to the Global Volcanism Program, the lava field contains several Holocene volcanic vents. The web site is telling that a boiling lava lake was also present in the mid 19th century [2].

When we observe this desert from space, we discover that this harsh environment was probably quite populated in ancient times. We can conclude this fact from some huge stone structures, the "desert kites", that can be easily seen in the images recorded by satellites. These structures were firstly observed by pilots of the Royal Air Force in the 1920s, flying over the desert. These pilots named them "kites", because these lines reminded of kites used by children to play, but in fact they are huge hunting traps [3].

We usually imagine our ancestors, before they settle down, as people simply hunting and gathering for food, but this is not true. The "desert kites" are the remains of an ancient hunting technique based on stone-walled traps, the construction of which surely involved several people for long times. The desert kites were used to push large herds of animals into some enclosures, or, in the worse case, to fall off from steep cliff edges [3]. The simplest structure of a desert kite has a triangular shape, consisting of two long, low walls built of stones and arranged in a V-shape, like a funnel, ending as a corral. Hunters pushed the game between the walls, trapping then the animals into the end of the structure. It is usually considered that animals were slaughtered "en masse" [3-5]. The faunal remains found in these sites are rare, and are including gazelle, Arabian oryx, and other species that are now rare or driven to extinction in the Levant [3]. A research at the Mesopotamian site of Tell Kuran, Syria, found what seems to be a deposit from a mass kill. According to [5], it was the overuse of desert kites a possible cause of extinction of several species. Typically, a desert kite possesses two, three or more small circular enclosures on the edge of its corral (in Fig.1 a wonderful structure as it can be observed with Google Maps, other examples in Fig.2 and 3). Some ancient rock art images show these hunting traps [7,8], depicting the role of the 'walls' of the kites. Let us note that these walls are low and then not able to stop any game. In fact, the walls are not walls at all: they are the basements, in the rocky harraat, where stick some poles and build a fence with branches. These structures create a visual effect as a barrier for the animals.

References 5 and 6 are telling that the kites were used for the slaughter of animals. There are also different opinions: for instance, Ref.[7] is telling that these hunting traps where not only designed to just capture and then kill animals, but also for the conservation of food, keeping the game alive, in small huts at the end of the kites. The kites were probably the first step from hunting to the intentional breeding of animals [7,8]. Other scholars do not agree with such conclusions [9].

It impresses the number of kites that we can still identify: it is seems to be in the thousands, distributed on the Arabian and Sinai peninsulas and, northward, as far as Turkey. Over a thousand have been recorded in Jordan alone [3]. The earliest desert kites are dated to the Pre-Pottery Neolithic B period of 9th-11th millennia BP. It is the radiocarbon analysis on charcoal within the kite pits helping to date the site [3]. As we discussed in the paper on Arabia [10], an interesting fact is necessary to note [7]. In the Khaybar area, Arabia, there are some remains of Neolithic villages very close to the hunting desert kites. Linking the "desert kites" with "human villages" could be a mistake, because wild animals are avoiding places where people live. Since the Khaybar area is full with ancient burial structures, sometimes placed inside the desert kites, a possible conclusion is that these burial mounds are more recent than the desert kites [7]. Because this conclusion is coming from the observation of satellite images, it seems that the satellites, and in particular the Google Maps, can help the archaeological researches providing a portrait of the human collective activities in the early stages of civilization.

On the Syrian Desert too, we see smaller circular stone structures and mounds in or near the desert kites (see Fig.2 for instance). We could then ask ourselves what are the remains in this desert of the human settlements at Neolithic times. For what concerns this question, interesting news was announced last year [11]. Robert Mason, an archaeologist with the Royal Ontario Museum, discovered near the Deir Mar Musa al-Habashi monastery, some tombs and stone circles, defining this site as a Syria's Stonehenge From stone tools found there, it's likely that the features date in the Middle East's Neolithic Period, roughly ranging from 8500 BC to 4300 BC. In Western Europe, the first structures built of stone date back approximately to 4500 BC. The Syrian site is then quite older than the European sites. Edward Banning of the University of Toronto says that more fieldwork is necessary because it is possible that the landscape that Robert Mason has identified could be an example, in the Neolithic period, of burial practices out of the settlement, that is, an off-site cemetery [11].

The report in [11] is continuing reporting a suggestion by Julian Siggers, of the Royal Ontario Museum: remembering that agriculture spread from the Near East to Europe, he is proposing the possibility that the stone landscapes, that is the creation of stone circles, have travelled with agriculture. On the other hand, Banning is replying that stone structures are found throughout the world and that people in Western Europe could have developed the stone landscapes independently of the people in Middle East. According to Edward Banning, the site studied by Mason is not unique [11,12]. "Archaeologists have detected, via satellite photos, what appear to be cairns and stone circles in other areas, including the deserts of Jordan and Israel. However, he admits that most of these things have not received a lot of archaeological investigation."

It is quite interesting what Ref.11 is telling, that the satellites are revealing so many structures, that probably, it is impossible to investigate all of them, and then many are not receiving specific investigation. We can by ourselves check the existence of stone circles in the Syrian Desert, using the Google Maps or Acme Mapper for instance. Here again, these map services are excellent to reveal the landscapes of the past. From Fig.4 to Fig.8, I am proposing some images of stone circles, sometimes with radial structures inside. There are also complexes composed by several stone structures (a collection of images at [13]).

To date the petroforms [14] that we see in the Syrian Desert, a huge fieldwork in the desert is necessary. Since this work needs time and financial support, the satellite imagery, as the Google Maps are clearly demonstrating, is a quite good source of information to locate the sites, propose their preservation and plan archaeological expeditions, with the support of contemporarily geophysical researches, which are surely performed in this area full of oil resources.

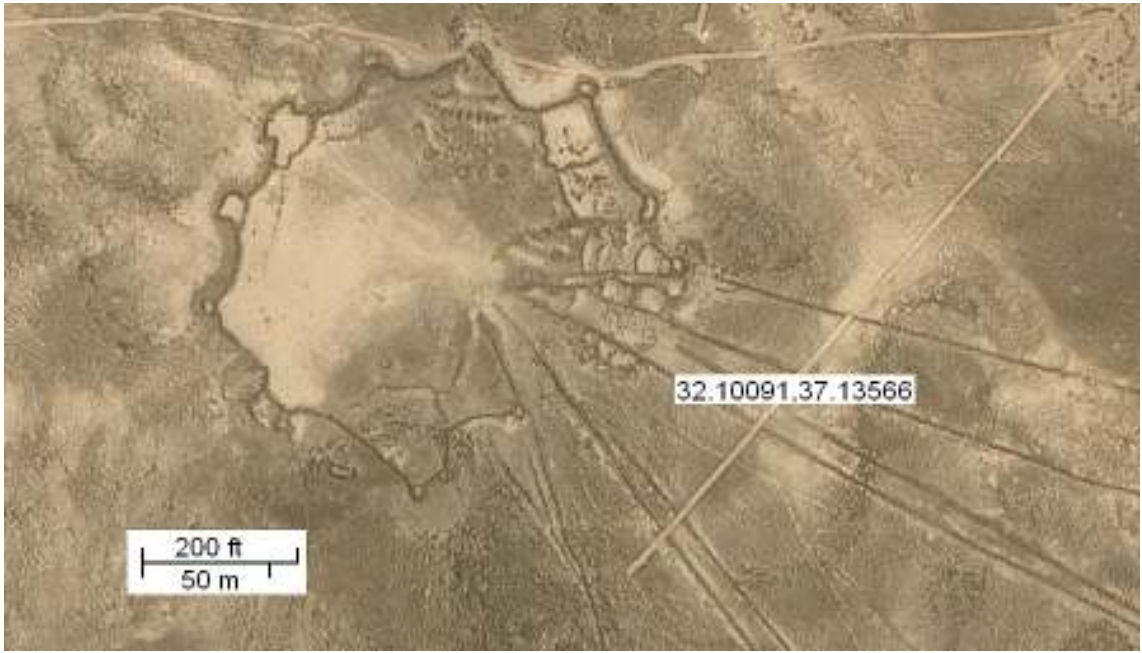

Fig.1. The wonderful structure of a "desert kite"(Jordan), as it can be observed with Google Maps.

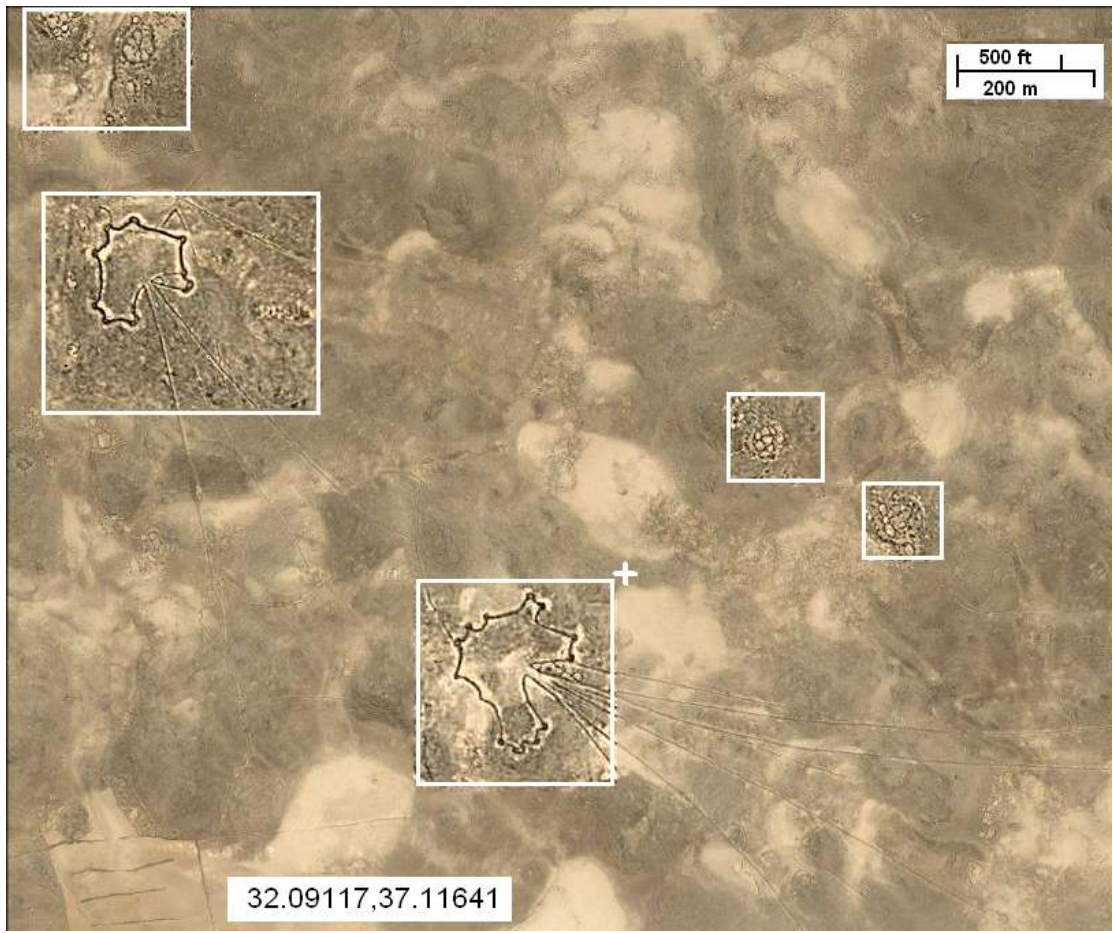

Fig.2: A landscape with desert kites and other stone structures (Jordan). This is an image adapted from the Google Maps. The visibility of marked areas has been adjusted with image processing software.

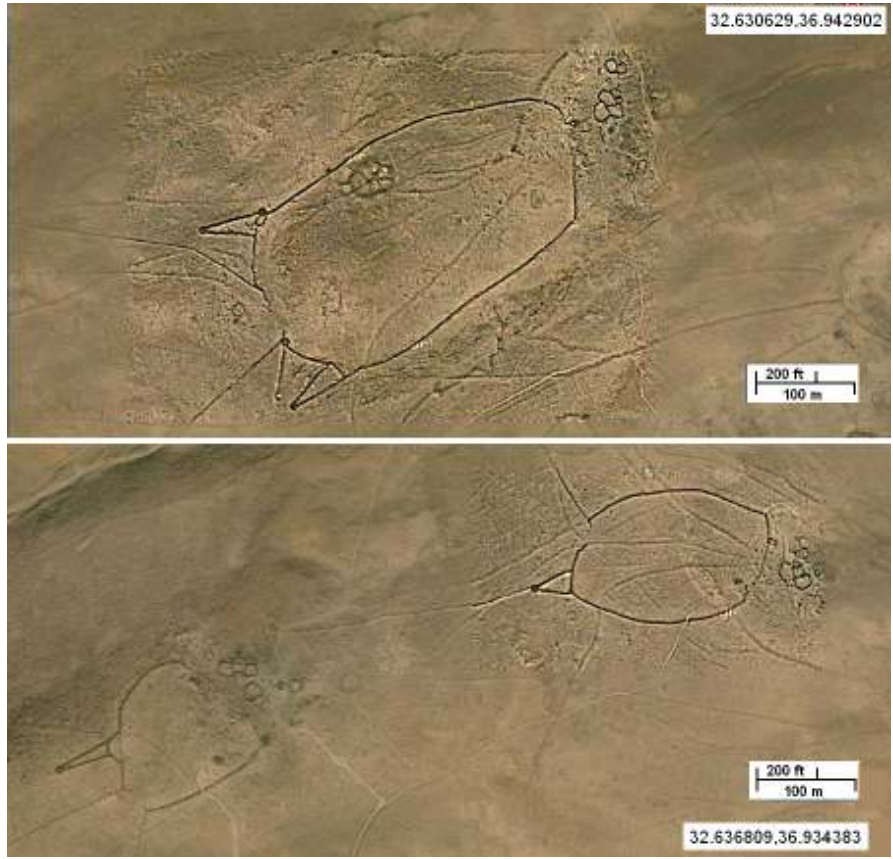

Fig.3: Desert kites. The images have been adapted from the Google Maps. The visibility of the stone strctures has been adjusted with image processing software.

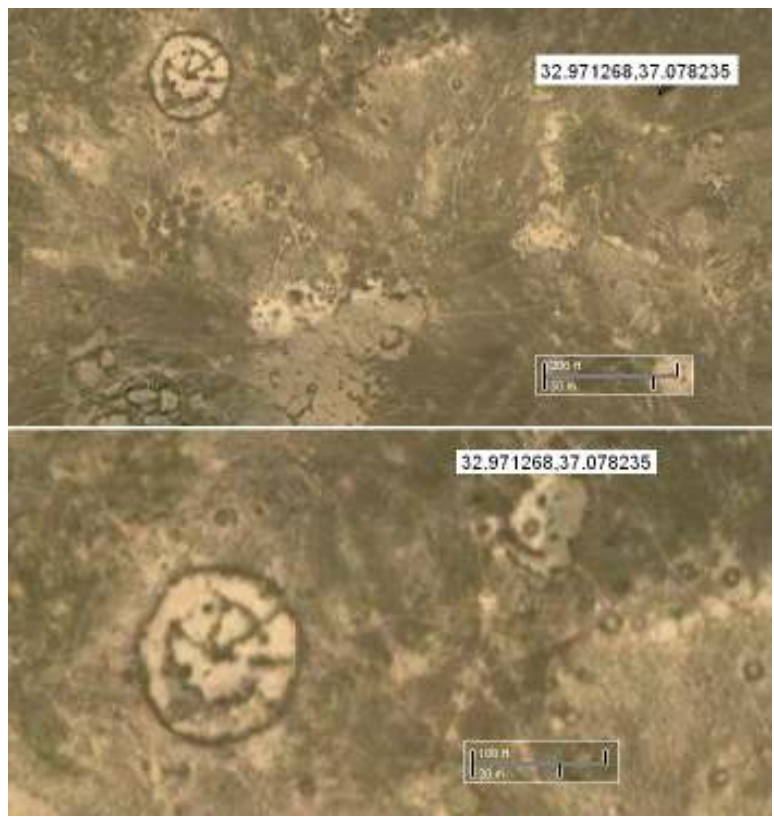

Fig. 4: A stone circles in Syrian Desert. The images have been adapted from the Google Maps.

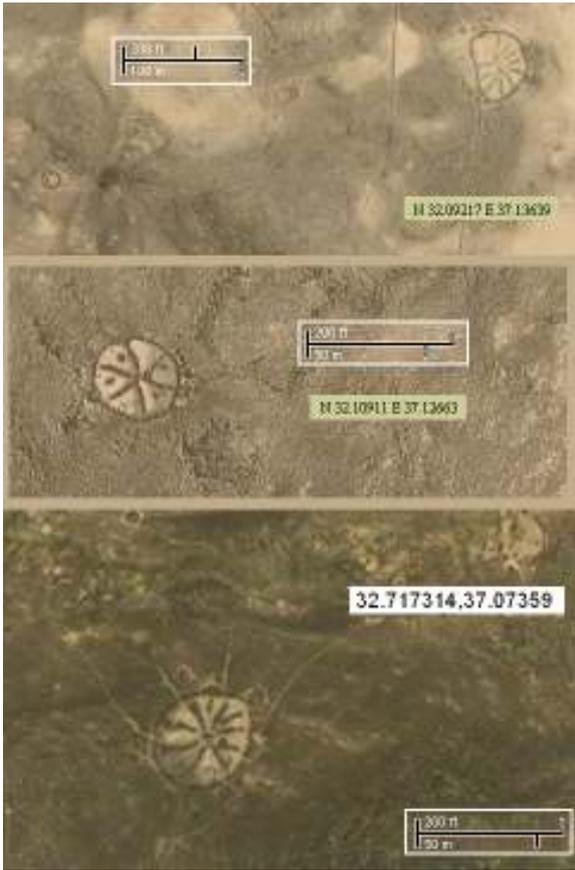

Fig. 5: Other stone circles in Syrian Desert. The images have been adapted from the Acme Mapper.

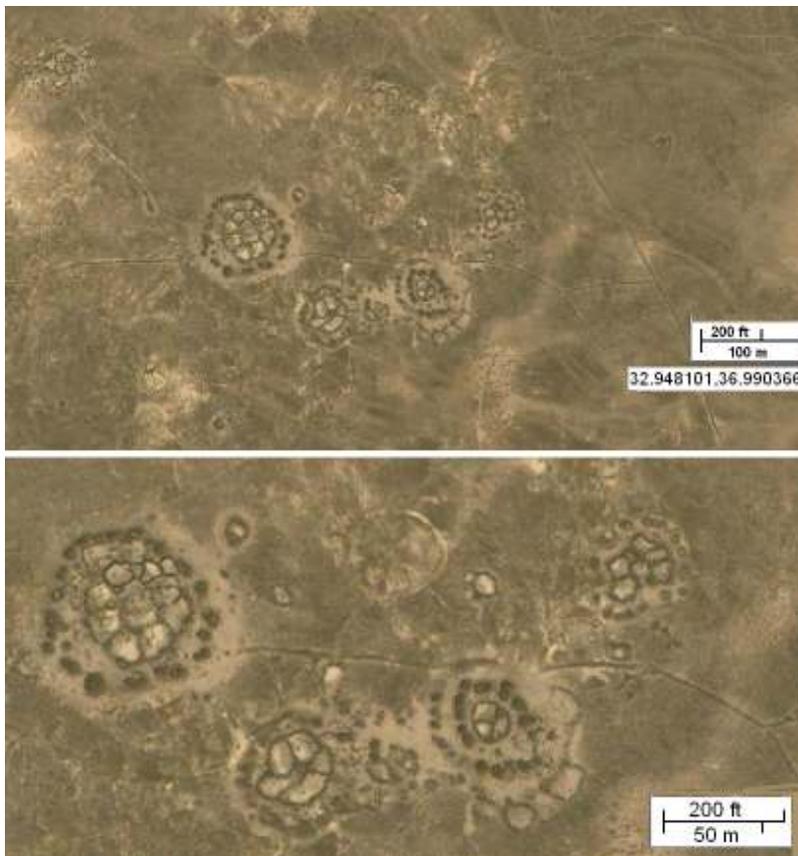

Fig. 6: Stone landscape in Syrian Desert. The images have been adapted from the Google Maps. Note the "dots" that are surrounding the circular complexes.

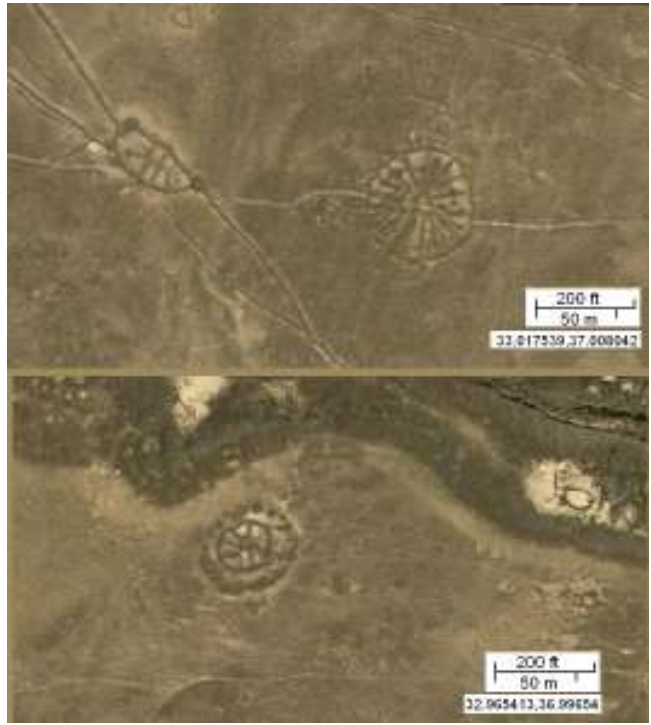

Fig. 7: Stone circles with radial structures. The images have been adapted from the Google Maps

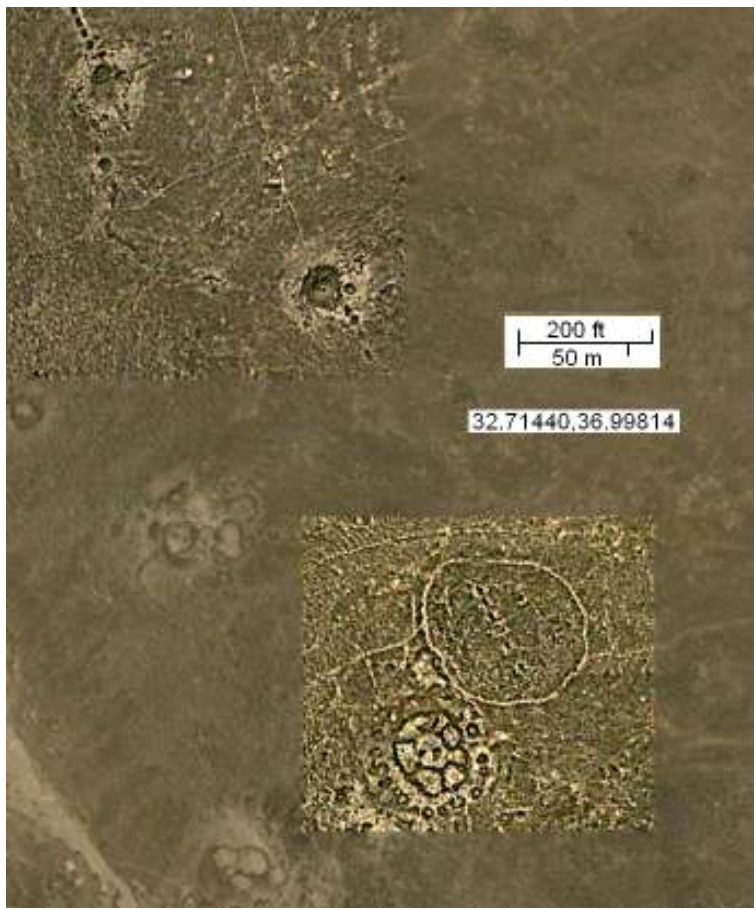

Fig.8: Mounds, lines of dots and circles in this stone landscape. The image has been adapted from the Google Maps. The visibility of the stone structures has been adjusted with image processing software.